\documentclass{IAU-TrB}
\usepackage{graphicx}
\usepackage{natbib}
\newcommand{\NOPRINT}[1]{\null}
\makeatletter
\let\jnl@style=\rm
\def\ref@jnl#1{{\jnl@style#1}}
\def\aj{\ref@jnl{AJ}}                   
\def\actaa{\ref@jnl{Acta Astron.}}      
\def\araa{\ref@jnl{ARA\&A}}             
\def\apj{\ref@jnl{ApJ}}                 
\def\apjl{\ref@jnl{ApJ}}                
\def\apjs{\ref@jnl{ApJS}}               
\def\ao{\ref@jnl{Appl.~Opt.}}           
\def\apss{\ref@jnl{Ap\&SS}}             
\def\aap{\ref@jnl{A\&A}}                
\def\aapr{\ref@jnl{A\&A~Rev.}}          
\def\aaps{\ref@jnl{A\&AS}}              
\def\azh{\ref@jnl{AZh}}                 
\def\baas{\ref@jnl{BAAS}}               
\def\bac{\ref@jnl{Bull. astr. Inst. Czechosl.}}
\def\caa{\ref@jnl{Chinese Astron. Astrophys.}}
\def\cjaa{\ref@jnl{Chinese J. Astron. Astrophys.}}
\def\icarus{\ref@jnl{Icarus}}           
\def\jcap{\ref@jnl{J. Cosmology Astropart. Phys.}}
\def\jrasc{\ref@jnl{JRASC}}             
\def\memras{\ref@jnl{MmRAS}}            
\def\mnras{\ref@jnl{MNRAS}}             
\def\na{\ref@jnl{New A}}                
\def\nar{\ref@jnl{New A Rev.}}          
\def\pra{\ref@jnl{Phys.~Rev.~A}}        
\def\prb{\ref@jnl{Phys.~Rev.~B}}        
\def\prc{\ref@jnl{Phys.~Rev.~C}}        
\def\prd{\ref@jnl{Phys.~Rev.~D}}        
\def\pre{\ref@jnl{Phys.~Rev.~E}}        
\def\prl{\ref@jnl{Phys.~Rev.~Lett.}}    
\def\pasa{\ref@jnl{PASA}}               
\def\pasp{\ref@jnl{PASP}}               
\def\pasj{\ref@jnl{PASJ}}               
\def\rmxaa{\ref@jnl{Rev. Mexicana Astron. Astrofis.}}%
\def\qjras{\ref@jnl{QJRAS}}             
\def\skytel{\ref@jnl{S\&T}}             
\def\solphys{\ref@jnl{Sol.~Phys.}}      
\def\sovast{\ref@jnl{Soviet~Ast.}}      
\def\ssr{\ref@jnl{Space~Sci.~Rev.}}     
\def\zap{\ref@jnl{ZAp}}                 
\def\nat{\ref@jnl{Nature}}              
\def\iaucirc{\ref@jnl{IAU~Circ.}}       
\def\aplett{\ref@jnl{Astrophys.~Lett.}} 
\def\apspr{\ref@jnl{Astrophys.~Space~Phys.~Res.}}
\def\bain{\ref@jnl{Bull.~Astron.~Inst.~Netherlands}} 
\def\fcp{\ref@jnl{Fund.~Cosmic~Phys.}}  
\def\gca{\ref@jnl{Geochim.~Cosmochim.~Acta}}   
\def\grl{\ref@jnl{Geophys.~Res.~Lett.}} 
\def\jcp{\ref@jnl{J.~Chem.~Phys.}}      
\def\jgr{\ref@jnl{J.~Geophys.~Res.}}    
\def\jqsrt{\ref@jnl{J.~Quant.~Spec.~Radiat.~Transf.}}
\def\memsai{\ref@jnl{Mem.~Soc.~Astron.~Italiana}}
\def\nphysa{\ref@jnl{Nucl.~Phys.~A}}   
\def\physrep{\ref@jnl{Phys.~Rep.}}   
\def\physscr{\ref@jnl{Phys.~Scr}}   
\def\planss{\ref@jnl{Planet.~Space~Sci.}}   
\def\procspie{\ref@jnl{Proc.~SPIE}}   

\makeatother

\title[STAR CLUSTERS AND ASSOCIATIONS]     
{}

\author[DIVISION V COMMISSION 37]   
{}

\pubyear{2015}
\volume{Volume XXIXA}
\pagerange{1--?}
\date{15 Sep 2015}
\jname{Transactions IAU, Volume XXIXA}
\editors{Thierry Montmerle, ed.}
\begin{document}

\maketitle

{\bf

\large
\begin{tabbing}
\hspace*{65mm}       \=                                              \kill
DIVISION V \\COMMISSION 37         \> STAR CLUSTERS AND ASSOCIATIONS                                     \\
                     \> {\it (STAR CLUSTERS AND ASSOCIATIONS)}                             \\
\end{tabbing}

\normalsize

\begin{tabbing}
\hspace*{65mm}       \=                                              \kill
PRESIDENT            \> Giovanni Carraro     \\
VICE-PRESIDENT       \> Richard de Grijs    \\
PAST PRESIDENT       \> Bruce Elmegreen                    \\
ORGANIZING COMMITTEE \> Peter Stetson, Barbara Anthony-Twarog, \\
                     \> Simon Goodwin, Douglas Geisler,\\ 
                     \> Dante Minniti\\
\end{tabbing}

\bigskip
\noindent
HIGHLIGHTS of COMMISSION 37 science results.
}

\small


\vspace{3ex}
\noindent 
\section{Abstract}
It is widely accepted that stars do not form in isolation but result from the fragmentation of molecular clouds, which in turn leads to star cluster formation. Over time, clusters dissolve or are destroyed by interactions with molecular clouds or tidal stripping, and their members become part of the general field population. Star clusters are thus among the basic building blocks of galaxies.
In turn, star cluster populations, from young associations and open clusters to old globulars, are powerful tracers of the formation, assembly, and evolutionary history of their parent galaxies.  Although their importance (e.g., in mapping out the Milky Way) had been recognised for decades, major progress in this area has only become possible in recent years, both for Galactic and extragalactic cluster populations. Star clusters are the observational foundation for stellar astrophysics and evolution, provide essential tracers of galactic structure, and are unique stellar dynamical environments. Star formation, stellar structure, stellar evolution, and stellar nucleosynthesis continue to benefit and improve tremendously from the study of these systems. Additionally, fundamental quantities such as the initial mass function can be successfully derived from modelling either the Hertzsprung$-$Russell diagrams or the integrated velocity structures of, respectively, resolved and unresolved clusters and cluster populations. Star cluster studies thus span the fields of Galactic and extragalactic astrophysics, while heavily affecting our detailed understanding of the process of star formation in dense environments.\\
This report highlights science results of the last decade in the major fields covered by IAU Commission 37: Star clusters 
and associations. Instead of focusing on the business meeting - the out-going president presentation can be found here:
{\bf www.sc.eso.org/~gcarraro/splinter2015.pdf} - this legacy report contains highlights of the most important scientific achievements
in the Commission science area, compiled by 5 well expert members.

\small


\vspace{3ex}
\noindent 
\section{Embedded and massive star clusters in the Milky Way: prepared by Ignacio Negueruela, Universidad de Alicante, Spain}
The past decade has been an era of discovery in the Milky Way, ushered in by our increasing capability to see through dust. This text is a short summary of the observations that have helped shape our understanding of massive star formation and young clusters in the Galaxy. I must apologise from the beginning for all the very relevant work that I have not been able to review. My reference list is just intended as a collection of examples to illustrate the broad range of excellent research carried out by our community. It is only natural that I concentrate on what I know best.

Deep infrared surveys of the Galactic Plane have revealed hundreds of new open clusters hidden by high obscuration. The publication of the 2MASS catalogue was followed by a number of dedicated searches, some by eye (e.g. \citealt{dutra03}), others using automated methods (e.g. \citealt{froebrich07}), that resulted in close to one thousand new cluster candidates. The {\it Spitzer}/GLIMPSE survey (\citealt{churchwell09}) presented us with a mid-IR view of the Galactic Plane at high spatial resolution, leading to a new collection of even more highly absorbed  clusters (e.g. \citealt{mercer05}). The process has continued as new, deeper surveys, such as VVV, have come on line (e.g. \citealt{borissova14}; \citealt{barba15}).

Many of these clusters are low-mass stellar groups still embedded in their parental clouds (e.g. \citealt{soares08}). Others are more massive young clusters with high-mass stars (e.g. \citealt{messineo09}; \citealt{chene13}). Finally, a significant number are simply old open clusters affected by interstellar extinction. Discriminating between these types is not always easy without more detailed investigation. Even though some semi-automated methods have been proposed to discern their nature (e.g. \citealt{beletsky}), follow-up deep photometry (e.g. \citealt{ivanov05}) and, above all, (in most cases, near-IR) spectroscopy is necessary for a good characterisation (e.g. \citealt{zhu09}; \citealt{messineo10}). Because of this, a large number of these candidates still remain unexplored.

Infrared imaging has also resulted in a novel picture of the interstellar medium. Large numbers of bubble-like structures have been found (\citealt{churchwell06}; \citealt{simpson12}), likely marking sites of massive star formation (\citealt{deharveng10}). Highly reddened high-mass stars have been identified in the vicinity of many them  (e.g. \citealt{marco11}; \citealt{pinheiro12}). In less obscured environments, bright-rimmed clouds also seem to mark sites of ongoing star formation in the immediate neightbourhood of high-mass stars (e.g. \citealt{panwar14}). {\em Herschel} observations have shown the concentration of proto-stellar cores along filaments. This increasing body of data supports a view of massive star formation as a process extending over large spatial scales, and thus resulting in measurable age spreads in many young regions (e.g. \citealt{bik12}). Infrared imaging has also allowed the identification of bow-shocks produced by high-mass stars ejected from massive clusters (e.g. \citealt{peri12}), giving rise to a lively theoretical discussion on the origin of isolated high-mass stars.

Adopting a more global point of view, statistical studies of the distribution of mid-IR sources have provided strong constraints on massive star formation in the Milky Way, providing estimates of the current global star-formation rate (e.g. \citealt{robitaille10}), typical duration of different phases (e.g. \citealt{mottram11}; \citealt{morales13}) or incidence of triggered star formation (e.g. \citealt{thompson12}). More recently, and in contrast, {\em Herschel} has provided a very detailed view of gas and dust around a few massive star-formation sites (e.g. \citealt{russeil13}).

But the opportunities for discovery do not only come from infrared imaging. The combination of X-ray imaging of young clusters with IR catalogues has become the preferred method to investigate the formation of low-mass stars in environments dominated by high-mass stars (e.g. \citealt{prisinzano11}; \citealt{feigelson13}), providing valuable input for  theories of high-mass star formation (e.g. \citealt{kuhn14}; \citealt{rivilla14}). Radio surveys have detected large numbers of new H\,{\sc ii} regions (e.g. \citealt{anderson11}), whose ionising stars or clusters remain in most cases still unknown. Radio observations have also produced one of the most significant advances in the past few years: the generalised use of geometric parallaxes to masers for  determination of distances to star-forming regions, an extremely powerful tool to probe the structure of the Milky Way (see references in \citealt{reid14}).

The current view of large-scale star formation in the Milky Way, emerging from all these developments, is complex. A typical configuration may be illustrated by the G305 star-forming region. High-mass stars in two moderately massive young clusters, Danks~1 and~2, drive a huge wind-blown bubble, whose rim is teeming with star-formation sites, where embedded lower-mass proto-clusters are forming just now. A diffuse population of proto-stars is scattered over the whole region (\citealt{hindson13} and references). The output of such a process is a classical OB association with massive central clusters, such as Per~OB1 or Cas OB8. Many giant molecular clouds seem to harbour similar configurations, among them the Carina Nebula (e.g. \citealt{preibisch11}), where Trumpler~14 has a mass $M_{{\rm cl}}\approx 10^{4}\:M_{\odot}$ (\cite{ascenso07}), or the W3 region (\citealt{roman15}) that seems to form a larger structure with W4 and W5, included within Cas~OB6. Others, such as W33 (\citealt{messineo15}) or W51, lack the central massive clusters, and will very likely evolve into dispersed associations, similar to Cyg~OB2 (\citealt{wright14}). This may also be the future of one of the most powerful star-forming regions known in the Milky Way, W49A (\citealt{homeier05}).

As a result of spatially concentrated star formation, we find massive young clusters. For many years, it was believed that the Milky Way lacked clusters with masses $\geq10^{4}\:M_{\odot}$. Then the obscured open cluster Westerlund~1 (Wd1) was found to host a population of $>70$ supergiants of spectral types ranging from O to M (\citealt{clark05}; \citealt{negueruela10}). A direct extrapolation of the number of massive stars detected would suggests a mass $M_{{\rm cl}}\approx 10^{5}\:M_{\odot}$ for a standard initial mass function(IMF). Direct star counts in the infrared give a lower limit of  $5\times 10^{4}\:M_{\odot}$ (\citealt{gennaro11}). The cluster seems to have formed monolithically in a single burst lasting less than 0.4~Ma (\citealt{kudryavtseva12}).

Many clusters have joined the list of massive clusters over the past decade. Some of them were already known, but their masses have been revised upwards. For example, Trumpler~14, mentioned above, or NGC~3603, for which \cite{harayama08} derive a mass in the range $M_{{\rm cl}}= 1$\,--\,$1.6\times 10^{4}\:M_{\odot}$ with indications of a top-heavy IMF. These mass estimates depend strongly on the distance adopted to the cluster and, hence, the extinction law. This problem becomes especially acute for the clusters close to the Galactic centre, affected by very heavy and variable reddening, with an extinction law that deviates strongly from the Galactic average (e.g. \citealt{nishiyama09}). For example, the mass of the Arches cluster is hotly debated. The extinction law adopted or even possible colour terms due to the very high reddening affect the luminosity function, which has to be later translated into a mass function. Masses ranging from $M_{{\rm cl}}<10^{4}\:M_{\odot}$ to several $10^{4}\:M_{\odot}$ have been found depending on assumptions (see references in \citealt{clarkson12}). But even much less reddened clusters are subject to such uncertainties. An extreme case is Westerlund~2, for which distances ranging from $\sim3$ to $\sim8$~kpc have been claimed (\citealt{carraro13}), because of very different estimates of the reddening law.

Besides W49A, the strongest candidate for the most massive star-forming region in the Milky Way is W43 (\citealt{bally10}; \citealt{nguyen11}), located close to the point where the Galactic bar joins the Scutum-Crux arm. The vigour of star formation in this area is likely related to this location, as there are indications of inflows from the spiral arm towards the densest regions (\cite{motte14}). Not far from W43, and at about the same distance from the Sun, several clusters rich in red supergiants have been found (e.g. \citealt{figer06}; \citealt{davies07}). These clusters have ages in the 10\,--\,20~Ma range, and estimated masses between~2 and $>5\times10^{4}\:M_{\odot}$. At least two large associations with $>10^{5}\:M_{\odot}$ have been found around them (\citealt{negueruela11, negueruela12}). The presence of so many extreme examples of present or recent star formation within a small span of the Galactic Plane ($\sim6^{\circ}$) seems to hint at a very privileged location. Further support for this interpretation would come from the detection of a similarly hefty complex at the opposite end of the Bar. Even though at least one massive cluster has been found in that general direction (\citealt{davies12}), the exact position where we should expect this starburst region is not certain. The young massive cluster vdBH~222, though originally thought to mark this far end of the bar (\citealt{marco14}), seems to be closer to us (and the Galactic Centre), suggesting that very large associations do not necessarily form at special places, and so hinting at the possibility that star formation in the inner Milky Way may be rather more vigorous than in the outer spiral arms.

Finding massive young clusters is not an end in itself. They not only inform us about star formation in violent environments, but also play a significant role as laboratories for the study of high-mass star evolution. With its huge population of evolved high-mass stars spanning all stages of evolution, Wd1 is the prime example. The different evolutionary paths of isolated and binary high-mass stars can be directly observed and documented within its co-eval population (\citealt{clark14} and references therein). Similarly, multi-epoch spectroscopic surveys of young open clusters have become fundamental tools to understand the physical properties of high-mass stars (\citealt{evans05, evans11}, and related references). As we move into the era of large spectroscopic surveys, this potential will be fullfilled. At the same time, ALMA is starting to look at the sites of massive star formation with unprecedented resolution. In the near future, more advanced instrumentation on the 10-m class telescopes, the launch of JWST and the advent of giant telescopes with adaptive-optics-assisted instrumentation offer the promise of not only an extended era of discovery, but also profound advances in our understanding of how star formation on large scales proceeds, and how high-mass stars are born, evolve and return to the interstellar medium the enriched material that will feed the future stellar generations.

\section{Open clusters in the Milky Way: prepared by Elena Glushkova, Sternberg Observatory, Moscow, Russia}
 Open star clusters (OCs) have always received
special interest, because by studying these objects, it is
possible to make suggestions on the structure and kinematics of
the Galactic disk and its dynamical and chemical evolution. However,
there are two major obstacles to achieve these goals: 1) the
sampling of OCs is complete only up to distances of 800 -- 1000
pc and 2) no unbiased, homogeneous collection of 
fundamental parameters are available in the literature.
Concentration of interstellar gas and dust toward the Galactic
plane makes the detection of new OCs difficult especially in the
visible range. To build a homogeneous catalog of OC parameters,
homogeneous observational data and reliable methods of
determination of the distances, ages, and reddening of the
clusters are required. Nonetheless, these obstacles are cleared
step by step in numerous studies, each considering few, sometimes
a dozen, clusters at once.

The situation dramatically changed in the beginning of the 2000s,
when large sky surveys became available. They triggered a new wave 
of interest to discover new, and investigate already 
known, open clusters. The greatest attention
was drawn by Two-Micron All Sky Survey ({\it 2MASS}, \citealt{skrutskie06}),
whose data were collected in three near-$IR$ bands.
Searches for clusters were conducted either visually, on images 
recorded in one of the bands, or from the Point Source Catalog using an
automated routine (\citealt{ivanov02}; \citealt{dutra03}; \citealt{bica03}; \citealt{kronberger06};
\citealt{froebrich07}; \citealt{koposov08}; \citealt{glushkova10};
\citealt{schmeja14}; \citealt{scholz15}). They found about
2000 candidates, visible in 
infrared only. One thousand cluster candidates more
were mined in other surveys: {\it DENIS} (\citealt{reyle02}),
{\it USNO-A2.0} (\cite{altdrake05}), {\it GLIMPSE} (\citealt{mercer05}; \citealt{zasowski13}),
{\it UKIDSS GPS} (\citealt{solin12}), {\it VVV} (\citealt{borissova11, 
borissova12, borissova14}; \citealt{solin14}), {\it WISE}
(\citealt{camargo15}). \cite{rubke15} started the program
of searching for MAssive Stars in Galactic Obscured MAssive
clusterS ({\it MASGOMAS}) and developed a new automatic tool,
which allows the identification of a large number of massive ( a few
1000 solar masses) cluster candidates from the {\it 2MASS} and
{\it VVV} catalogs.

As {\it 2MASS} is a homogeneous and all-sky catalog, many attempts
were made to derive a homogeneous list of accurate physical
parameter for OCs out of  $JHK_S$ data: refer, for example, to the
papers by \cite{koposov08}, \cite{tadross08, tadross09, tadross11},
\cite{glushkova10}, \cite{bukowiecki11},
\cite{kharchenko13}. However, the
comparison of parameters obtained in these investigations with
respect to the largest catalog by \cite{kharchenko13}, gives
mean standard deviations for the distance, age and reddening as
1.2 kpc, 0.5 dex, and 0.27 mag, respectively (\citealt{netopil15},
which are definitely insufficient for comprehensive studies
of the structure of the Galactic disc. Involving data from large
surveys, scientists undertook massive determinations of other OC
characteristics: the radial velocities of 110 OCs and the
metallicities of 89 OCs were derived using data from RAdial
Velocity Experiment ({\it RAVE}, \citealt{conrad14}), the mean
proper motions and stellar membership probabilities for 1805 open
clusters were found using {\it UCAC4} (\citealt{dias14}). \\

Since open clusters provide  information about the chemical 
pattern of Milky Way disk, the study of their chemical composition 
attracts special interest. \cite{paunzen10} compiled data from the 
available sources in the literature to make up the catalog of OC 
metallicities based on photometric data, which lists 188 clusters 
(\citealt{paunzen10}) and set of high-quality cluster metallicities 
based on high-resolution
spectroscopic studies for 78 clusters (\citealt{heiter14}). \citealt{heiter14}
show that none of the current models are able to
satisfactorily describe the OC's metallicity as function of 
galactocentric distance. The present high-resolution spectroscopic 
Galactic surveys include OCs among their targets: Apache Point Observatory
Galactic Evolution Experiment ({\it APOGEE}, 
the Gaia-ESO Survey ({\it GES}, \citealt{gilmore12}), the GALactic 
Archaeology with {\it HERMES} ({\it GALAH}, \citealt{anguiano14}) supply accurate 
radial velocities and detailed chemical abundances. Using 100 OCs
from the uniformly observed complete {\it SDSS-III/APOGEE-1} 
DR12 dataset ({\it OCCAM} survey), \citealt{frinchaboy15} presented 
age and multi-element abundance gradients for the disk of Milky Way. 
The INfrared Survey of Young Nebulous Clusters ({\it IN-SYNC}, 
\citealt{covey15}) leverages the stability and multiplex capability 
of the {\it APOGEE} spectrograph to obtain high resolution spectra 
at near-infrared wavelengths and to study the dynamics and star 
formation history of young clusters. Four young stellar groups in Perseus 
and Orion molecular complexes were investigated (\citealt{covey15}) 
based on homogeneous stellar parameters derived from {\it APOGEE} 
spectra for thousands of pre-main sequence stars (\citealt{cottaar14}).
OCs are also included in the list for the space missions {\it Gaia} and {\it Kepler}. \\
 
Some projects were delivered especially to investigate OCs
 and derive their physical parameters in a precise and
homogeneous way: for example, within the framework of Bologna Open Clusters
Chemical Evolution project ({\it BOCCE}, \citealt{bragaglia06})
about 50 clusters were studied. {\it BOCCE} uses both
comparison between observed color-magnitude diagrams (CMDs)
and stellar evolutionary models, and the analysis of 
high-resolution spectra to derive age, distance and 
chemical composition. The Open Clusters Chemical Abundances from 
Spanish Observatories survey ({\it OCCASO}, \citealt{altcasamiquela14}) 
plans to derive abundances for more than 20 chemical species in at 
least 6 Red Clump stars in 30 northern OCs. Nine clusters have been studied by {\it OCCASO}.
Sejong Open Cluster Survey (SOS, \citealt{sung13}) -- a project dedicated 
to providing homogeneous photometry of a large number of OCs 
in the SAAO Johnson Cousins' $UBVI$ system,  has many relatively small 
sparse unstudied clusters among its targets. The {\it OPD} survey 
(\citealt{caetano15}) is {\it UBVRI} photometric survey of southern open OCs,
which is complementary to {\it OAN-SPM UBVRI} survey of 406
northern clusters (Michel et al. (2016)). 
Lasting 17 years, {\it WIYN} Open Cluster Study ({\it WOCS}, \citealt{mathieu13}) 
is dedicated to comprehensive photometric, astrometric, and
spectroscopic studies of select OCs. In their recent investigation 
(\citealt{thompson14}) new deep wide-field optical and near-infrared 
photometry ($UBVRJHK_S$) of the cluster M35 is presented, against which 
several isochrone systems are compared: Padova, PARSEC, Dartmouth, and $Y^2$. 
Two different atmosphere models are applied to each isochrone: ATLAS9 and BT-Setti. 
For any isochrone set and atmosphere model, observed data are accurately 
reproduced for all stars more massive than $0.7\:M_{\odot}$. For 
less massive stars, Padova and PARSEC isochrones consistently produce higher 
temperature than observed. Dartmouth and $Y^2$ isochrones with BT-Setti 
atmospheres reproduce optical data accurately; however, they appear too blue 
in IR colors. {\it WIYN's} extension -- the Southern Open Cluster Study ({\it SOCS}, 
\citealt{kinemuchi10}) includes 24 clusters, wide-field photometry results are
already available for three of them. 
\cite{faria14}
determined the radial velocity of stars belonging to a group of
open clusters using spectra with  spectral resolution of 4000 and plan
to calculate the mean radial velocities of a number of  OCs. To
investigate star formation processes, \cite{lim15} initiated
a photometric survey of young open clusters in the Galaxy and
already studied 13 famous OCs having a wide range of surface
densities (log($\sigma$)=-1 -- 3 $stars/pc^2$) and total 
masses ($500 \-- 50000\:M_{\odot}$) and also distributed in five different
spiral arms in the Galaxy. \cite{lim15} found that the slope
of the IMFs in the high-mass regime appears to be shallow for
massive compact clusters, and the mass of the most massive star in
a given cluster also has a tendency to be large in massive
clusters.
\cite{costa15} started a program that determines the properties 
of Local (Orion) spiral arm. They plan to carry on a comprehensive study 
of 29 young OCs which includes a $UBVRI$ photometric analysis 
and determination of their kinematics. The first cluster NGC 2302 has already been investigated. \\

When large catalogs and data sets became available in the
literature, and massive determination of parameters of clusters
became possible, numerous automated and semi-automated techniques
were developed to retrieve these parameters primarily by using
color-magnitude diagrams. One of the most powerful tools
is the  Automated Stellar Cluster Analysis package ({\it ASteCA},
\citealt{perren15}, which makes use of positional and photometric
data to provide accurate estimates of the cluster's metallicity,
age, extinction and distance values, and robust stellar cluster
image and photometry simulation package {\it MASSCLEAN} (\citealt{popescu10}), 
which creates synthetic clusters and generate CMD
templates for a variety of cluster masses and ages, and which
mimic the observational photometric errors when using isochrone
fitting (\citealt{popescu14}).\\

Thanks to All Sky Automated Survey ({\it ASAS}, \citealt{pojmanski05}),
a large number of new Galactic Cepheids have been
discovered during the last dozen years. That is why new attempts were
undertaken to find Cepheids attributed to open star clusters.
Taking into account all possible characteristics of these variable
stars and OCs, \cite{anderson13} found five new genetic
relations between Cepheids and clusters, and \cite{chen15}
reported 8 new Chepheid-cluster pairs. Some papers were devoted to
photometric and spectroscopic observations of known clusters
hosting the Cepheids, in order to confirm membership of Cepheids in
OCs and refine cluster's physical parameters (\citealt{majaess13a,
majaess13b, majaess12}, \citealt{turner12}).\\

One of the most interesting results of the studies of individual
clusters was published by \cite{davies11} on GLIMPSE-C01
referred to in the literature as an old globular cluster
traversing the Galactic disk. The authors obtained high-resolution
near-infrared spectroscopy of over 50 stars in the cluster and
found the average radial velocity is consistent with being part of
the disk, and determined the cluster's dynamical mass to be
$8\times 10^{4}\:M_{\odot}$. From analysis of the cluster's $M/L$ ratio
and location of the red clump, \cite{davies11} suggested the
cluster's age to be 400--800 Myr and concluded that GLIMPSE-C01 is
the most massive Galactic intermediate-age cluster discovered to
date.\\

This way, the number of known Milky Way open clusters increased 
from about 1500 to almost 4000 during the last 10 -- 15 years. The main 
physical parameters were derived for  most of them but quality of these 
determinations is insufficient to study the Galactic disk comprehensively.
That is why a lot of new different surveys were started to measure a large range 
of open cluster's properties.\\

\section{Globular clusters in the Milky Way: prepared by Angela Bragaglia and Eugenio Carretta, Osservatorio di Bologna,  Italy}

The last ten years have seen a renewed interest in Galactic globular clusters
(GC), mainly because they have been demonstrated to be much more complex and
intriguing than believed in the past. The main reason is that spectroscopic and,
later on, photometric observations have driven a dramatic shift  from
considering GCs as the best approximation of simple stellar populations (see
e.g. the review  by \citealt{rffp88}) to the simplest example of $multiple$
stellar populations (e.g. the review by \citealt{gratton04}).  We use here the
words ``populations" and ``generations" as synonym, implying that in the same GC
stars of (slightly) different age coexist. This is the generally accepted
scenario, although there are still many problems in explaining the mechanism of
GC formation and internal self-enrichment (in light elements for all GCs, in
heavier elements only for a fraction). For recent reviews see for instance
\cite{gratton12} for spectroscopic results, \cite{piotto09} for photometry, and
\cite{charbonnel15} for theoretical challenges. \\

We present here selected highlights in Galactic GC work of the last decade, with
a strong bias towards multiple populations and observations. \\

The abundance of light elements in GC stars shows large star-to-star scatter, at
variance with what happens for the bulk of field halo stars; these variations
are anti-correlated (C and N, O and Na, and Mg and Al are depleted and enhanced,
respectively). These so called ``anomalies" in light elements had already been
detected in many GCs, but generally only in giant stars. However,  the presence
of products of hot H-burning also in main sequence stars called for more massive
stars as original polluters (see subsection on models) and we now speak of
first-generation (FG) and second-generation (SG) stars in GCs. After the first
pioneering works, light element (anti-)correlations were routinely found also
among main sequence and unevolved stars, using also high-resolution spectra, see
e.g. \cite{carretta05,cohen05cn,kayser06,kayser08,pancino10,smolinski11,lardo12}
for C, N, and 
\cite{pasquini05,lind09,bragaglia10,dorazi10,monaco12,dobrovolskas14} for Na, O,
Mg, Al, and Li. 
This decade saw a wealth of observations on large samples of stars and of
clusters, also thanks to multi-object high-resolution spectrographs, like Hydra
or FLAMES, so that a quantitative analysis of the light elements
anti-correlation became possible.  Large scale studies were conducted, see the
long list of GCs and papers in \citealt{gratton12}, among which we have for
instance the closest GCs, i.e., M 4, NGC 6752, NGC 6397, and M 22
\citep[e.g.][]{marino08,carretta09a,carretta09b,yong08,korn07,marino11}, very
massive clusters like $\omega$~Cen \citep{marinoomega} and M 54
\cite{carrettam54}, and low mass GCs like NGC6838 \citep{cordero15}.  Recently,
also high-resolution surveys produced results in this field \citep{apogee,ges}.
All evolutionary phases were targeted. The red giant branch (RGB) was the
favourite, but also the horizontal branch (HB) was observed and in some GCs also
He was measured \citep[see e.g.][]{villanova,marino14,gratton15}, an important
diagnostics of the multipopulation scenario. Helium has also been measured in
RGB stars, but only in NGC 2808 \citep{pasquini11} and $\omega$~Cen
\citep{dupree13}.  More recently, the Asymptotic Giant Branch (AGB) observations
showed that  at least in some GCs the AGB stars do not show the same level of
modification in light element abundances (\citealt{campbell,charbonnel13}, but
see also \citealt{johnson15}).
Briefly, in all the examined MW GCs  (and interestingly, also in GCs of the LMC 
(\citealt{johnson06,mucciarelli09} and Fornax \citealt{letarte06}), a prominent
Na-O anti-correlation has been found. Possible, rare, exceptions are very
low-mass GCs, like the two Sgr clusters Ter 7 and Pal 12 \citep[e.g.][]{ter7}, and
Rup106 \citep{villanova13}. The Na-O anti-correlation appears almost a defining
properties of (massive) GCs and is not present among field
stars or open clusters (\citealt{bragaglia12,bragaglia14,cunha,maclean15}, but
see \citealt{geisler12} for a different opinion on NGC6791).
An even more extreme departure from the simple stellar population paradigma
comes from the finding of GCs with dispersion in iron and heavy elements. After
$\omega$~Cen, also M 22 was found to display an intrinsic metallicity spread by
\cite{dacosta09,marino09} (recently challenged by \citealt{mucciarellim22}),
correlated with a spread in neutron-capture elements. Among  GCs displaying iron
and n-capture element spreads are for instance M 54, NGC 1851, and M 22.\\

The impact of different abundances of light element ions the photometric
properties of stars in GCs (especially in the filters containing molecular
features of CNO elements, see \citealt[e.g.][]{sbordone11}) is exploited to
better understand the origin of GCs. Different abundances between FG and SG
stars translate into spread and even split sequences along the whole CMD of GCs.
These multiple sequences are observed with several photometric systems,
including broad band \citep[e.g.][]{marino08,milone08,han09,lardo}, intermediate
band \citep{yong08,carretta11},  and narrow band  \citep{lee,lim}.  The recent
exploitation of the UV Hubble Space telescope (HST) filters  allows to reach also features of OH
hydride (see the UV GC survey described in \citealt{p8}).  HST observations
reveal split sequences all the way from the main sequence (MS) up to the red giant branch (RGB) and
horizontal brand (HB, see for instance the spectacular main sequence (MS) of NGC 2808, \citealt{piotto07}).   The
large photometric samples allow to study the radial distribution of stellar
populations across the cluster area \citep[e.g.][]{lardo, milone6752,kravtsov}. 
Usually  SG stars are found more centrally concentrated, as predicted by most
scenarios of GC formation. Photometry 
permits to detect even discrete populations in a single GC, like NGC 2808
\citep{milone2808}, confirmed by abundance analysis of individual stars
\citep{carretta2808}.
The coupling between precise spectroscopy and photometry unravels a variegated
landscape for Galactic GCs, that come in different flavours. Most GCs are
monometallic (concerning iron and heavier elements). A growing number of objects
shows a spread in Fe (see above).  They also share a few other common
properties: a correlation between Ca and Fe (suggesting enrichment by type II
SNe), and an enhancement of elements from slow neutron-capture process observed
among stars of the more metal-rich cluster component. This suggests that, like
$\omega$Cen and M 54, the remnant nuclei of former dwarf galaxy 
\citep[e.g.][]{bf03,bellazzini}, these GCs may be the final products of cluster 
formation in a dwarf galaxy environment \citep[e.g.][]{bk11}.
Maybe related to the multiple populations in GCs, in the LMC and SMC many
intermediate-age clusters with extended or even split main sequence turn-off's
and red clumps have been found \citep[e.g.][]{milone09,girardi09,correnti15}.
This may be due to extended star formation \citep[e.g.][]{goud11} or to stellar
rotation \citep[e.g.][]{bdm09} or binarity \citep[e.g.][]{yang}. No light
elements anti-correlation has yet been found in these clusters
\citep{mucciarellilmc}.

Observational evidence of multiple stellar populations calls for some FG stars
to have polluted material from which the SG stars formed. The most commonly
discussed FG polluters are intermediate mass  asymptotic giant branch (AGB) stars \citep[][e.g.]{ventura08}
and fast rotating massive stars  \citep[FRMS][e.g.]{decressin07}, but also
interactive massive binaries have been proposed \citep{demink09}, very massive
stars \citep{ds14}, or early disk accretion \citep{bastian13}. However, multiple
populations also pose tight constraints that challenge (m)any model(s) of GC
formation. The currently observed ratio of FG to SG stars (about 1/3 and 2/3,
\citealt{carretta10,bl15}) is not easily reconciled with the amount of yields
provided by any  candidate FG polluter (see for instance \citealt{bastian15}). 
This ``mass budget" problem was often circumvented by scenarios assuming that
GCs were initially from 10 up to 100 times more massive than present-day GCs
\citep{bekki07,decressin07,dercole08,dercole10,carretta10,sc11}. However,
evidence is growing that in external dwarf galaxies harbouring old GCs (showing
multiple populations)  these objects cannot have been more than 3-4 times more
massive at their formation epoch, since they already account for about 25$\%$ of
the galaxy mass in metal-poor stars \citep{larsen12,larsen14,tudorica15}. 
Another severe challenge is posed by the very same existence of the observed
anti-correlations, which require  a certain amount of pristine gas to be mixed
with ejecta of polluters to reproduce the observations. In some scenario (e.g.
with AGBs, \citealt{dercole10}) dilution is mandatory since AGBs produce a 
correlation between Na and O abundance. From where this gas with primordial
composition came after the GCs were swept by the first type II Supernovae is an
issue still open, widely discussed but with no completely satisfactory answer 
up to now.  A recent scenario proposed  by \citealt{trenti15} shows a possible
way out, with blobs of pristine gas  nearby the outcome of a major mini-halo
mini-halo merger at high redshift, which  could be later accreted refurbishing
the evolving proto-GC with fresh reservoir of diluting matter. This mechanism
could also be an attractive way to explain the discreteness observed both in
CMDs and in the anti-correlations (see e.g. \citealt{milone6752, milone2808} and
\citealt{carretta12,carretta2808}, for NGC 6752 and NGC 2808).\\

Clusters lose mass and stars during their evolution (see, e.g.,
\citealt{baumgardt08}). Stars probably lost by GCs have been searched for in
halo samples using the peculiar SG chemistry. CN band strength was used by
\cite{mg10,martell11} on Segue SDSS spectra, while enhanced Na and/or depleted O
abundances were considered by \cite{carretta10,ramirez12}. These papers provided
a fraction of SG-like halo stars of about 1.5 to 3$\%$.  Recently, using data from
the public spectroscopic survey Gaia-ESO \cite{lind15} found a probable GC
escapee. For a discussion on the contribution of GC stars to the halo, see for
instance \cite{gratton12}.
With the advent of large photometric surveys such as SDSS, Pan-STARRS, DES, VST
ATLAS, VVV, etc new GCs have been found, see e.g.,
\cite{laevens15,belokurov14}, even if the objects found are
sometimes classified as GCs or ultrafaint dwarfs depending on the study. The
Galactic GC populations has increased in number, but the newly discovered
clusters are all low-mass systems.
Wide field imaging has been used systematically to look for extended and
extra-tidal structures \citep[e.g.][]{jordi10,carballo} In a few cases, tidal
tails or streams still connected to the originating GC have been found   (e.g.,
NGC 288, \citealt{grillmair13} or Pal 14, \citealt{sollima11}), but there are
many more without a clearly associated progenitor  cluster.
Furthermore, streams, moving groups, common proper motions groups have been
sometimes  associated to now dissolved clusters, in some cases using chemical
tagging. Examples of a positive  and negative identification, respectively, are
the Aquarius stream (consistent with being GC debris, \citealt{wdb12}) and the
Kapteyn group (not associated with $\omega$~Cen, \citealt{navarrete15}).\\

``Exotic'' objects in GCs comprise for instance blue stragglers stars (BSS), 
low-mass X-rays binaries, and millisecond pulsars (MSP), all very good tracers
of the  evolution of close binary systems in dense environments and of the
dynamical history of the parent cluster.
The study of  BSS proceeded with the collection of large samples covering the
entire extension of the GCs \citep[e.g.][]{dalessandro,knigge,salinas}, in the
determination of their dynamical status
\citep[e.g.][]{ferrarom30,simunovic,ferraro12} and chemical composition
\citep[e.g.][]{lovisi}. For more detailed and recent highlights on BSS, see for
instance the contributions in `Ecology of Blue Straggler Stars', held in 2012.
\cite{ransom} found 21 new MSP in tbe massive GC Terzan 5, which shows the
largest number of X-ray sources among GCs. This cluster is particularly
interesting also because \cite{ferraroter5} found it hosts two stellar
populations with different iron contents and ages and proposed it to be the
remnant of one of the primordial building blocks that formed the bulge (and not
a true GC). The different metallicity was later confirmed by
\cite{origlia,massari}.

\section{Extra-galactic star clusters: prepared by Tom Richtler, Universidad de Concepci\'on, Chile }
The following pages try to extract the results from 10 years of world wide intensive research using the extremely short style of a conference summary and applying a strong bias to observational work. 
\cite{brodie06} review  "Globular clusters and Galaxy Formation", \citet{harris10} "massive star clusters in galaxies",  \citet{portegies10} "young massive clusters" , \citet{kruijssen14}  concepts of GC formation.\\

New M31 GCs have been detected, many within the "Pan-Andromeda Archaeological Survey"  \citep{mcconnachie09} and the data base for M31 GCs has been significantly increased both in quantity and quality \citep{puzia05,kim07,lee08,fan10,huxor14,ditullio13,ditullio15}. 
A updated analysis of the properties of the M31 GCS is given by  \citet{huxor11}.  A flattening of the surface density profile at a radius of about 30 kpc corresponds to a
flattening in the stellar surface brightness profile, which might indicate an accretion of the outer halo. More direct evidence for accreted clusters is found in \citet{mackey10,mackey13}.
Structural parameters  are presented by \cite{barmby07,peacock10}. 300 metallicities of old GCs have been derived by  \citep{caldwell11}. The metallicity distribution is unimodal in contrast to
that of the Milky Way and giant ellipticals.  Detailed element abundances for  GCs from integrated light have been derived by \citet{colucci09}.
\cite{strader11} combine kinematic data and structure parameters to derive  {\it M/L}-ratios for 200 GCs and confirmed previous findings that the {\it M/L$_V$}-values decline with increasing metallicity,
contrary to what naively is expected from  stellar models. Shallower mass functions of metal-rich clusters can explain this. {\it M/L$_V$}-values also increase with cluster mass, possibly as a consequence of mass segregation. This is disputed by \citet{shanahan15}.
The kinematics of the outer halo GCs in M31 is studied by \citet{veljanoski14}. Groups of GCs are related to the debris of stellar streams.
The Fornax dwarf spheroidal  shows an  extremely large ratio of stars in GC to field  stars of a similar low metallicity. This constrains the loss of first generation stars in GCs \cite{larsen12}. \\

Nearby spirals with high star formation rates have been surveyed for clusters by \citet{bastian12} (M83), 
\citet{larsen11} used HST data to measure colour magnitude diagrams for resolved young massive clusters in nearby  spiral galaxies. There is no gap visible between the  H-burning main sequence
stars and the He-burning supergiants like in canonical isochrones. Age spreads of a few Myrs are able to fill this gap, may be also interacting binaries.  \\

The Initial Cluster Mass Function (ICMF) of young clusters in spiral galaxies has been investigated by \citet{larsen09}. A original Schechter function with a cut-off mass of $\approx 2\times10^5 M_\odot$
describes the GC luminosity functions well, if the luminosity evolution is only secular. \citep{elmegreen06}  provides insight into the similarity of cluster and stellar IMFs. 
Based on a sample of 37 nearby dwarf galaxies, \citet{cook12} study the relation between SFR and cluster formation. 
Galaxy mergers can host a plethora of  massive clusters. 
The Antennae galaxies are  prominent targets with  age distribution of GCs studied by \citealt{fall05}, new spectroscopic data \citep{whitmore05,bastian09},
general demographic model and application to the antennae \citep{whitmore07}. The latter work demonstrates that the enhanced number of massive clusters in mergers is a sample size effect due to the high star
formation rate.  The Antennae are also a test ground for the evolution of clusters, in particular the processes of disruption \citep{fall09,renaud08, karl11,renaud13}.  Some young clusters in the Antennae have been observed with ALMA, but large reservoirs of molecular gas in the clusters have not been found \citep{cabrera15}.  Cluster formation and disruption in mergers through simulations have been investigated by
\citet{kruijssen12}.  \citep{bastian13} give structural parameters for 36 clusters in NGC 7572.  Among them is W3, the most massive cluster known, whose profile extends out to 500 pc.  \\

The nearest merger remnant (and giant elliptical) is NGC 5128 (CenA).
Ages, metallicities are known for about 400 objects   \citep{gomez06,rejkuba07,woodley10a,woodley10}.   One third of this sample
show ages less than 8 Gyr.  \citet{taylor10} find  an increase of the M/L-values with the dynamical mass. 
 The GCS has been searched  over an area of 1.5deg$^2$   \citep{harrisg12}.   More than 1000 GCs are now known.  
NGC1316 (Fornax A) is after CenA the closest merger remnant. Its GCS host clusters of a wide age-range down to 0.5 Gyr (\citealt{richtler14}). Among them are objects as massive as $1.6\times10^7  M_\odot$
(\citealt{bastian06}). There is an isolated young star cluster complex which demonstrates GC formation outside periods of high star formation rate \citep{richtler12}.\\

The nearby galaxy clusters Virgo and Fornax have been intensively surveyed with HST/ACS which produced a bulk of papers. From the Virgo survey: \citet{peng08} study  formation efficiencies of GCs. The GC
mass fraction is lowest at intermediate host luminosity, the specific frequency is dominated by blue clusters. Dwarf galaxies near to M87 seem to be tidally stripped of their GCs.
 \citet{villegas10} on the GC luminosity function and distance determination: the dispersion of the GCLF correlated tightly with the host's absolute magnitude.
\citet{mieske06} on colour-magnitude relations in GCSs, \citet{liu11} on colour gradients within GCSs. 
\citet{georgiev10} find that the total mass  in GCs scales with the halo mass of the host galaxy. \\

A new catalogue by \citet{harris13} updates our current knowledge of GCS properties and relations with host galaxy properties. The number of GCs  obeys a  fundamental plane-like relation $N_{GC} \sim (R_e \sigma_e)^{1.3}$ 
for galaxies of all luminosities.   On the other hand the ratio mass  in GCs/halo mass is essentially constant
\citep{hudson14}, the GC number also scales with halo dark mass and \citet{harris15} even find a strict proportionality between the number of blue clusters and the halo mass. 
\citet{harris14} compare the GC luminosity functions of seven brightest cluster galaxies and find identical shapes.\\

Dwarf galaxies show a tendency of increasing specific frequency with decreasing luminosity \citep{miller07}.  
How the GCSs of dwarf galaxies are affected by galaxy harassment in clusters, is studied by \citet{smith13}. The critical parameter is the dark matter fraction that remains after interaction
processes. \citet{bruens11} explain both compact and extended clusters by dynamical evolution in star cluster complexes.
Massive young GCs are also seen in star forming  dwarf galaxies  \citep{adamo11} which in the context of hierarchical clustering may have been important for the assembly of GCSs.\\

The richest GCs of nearby galaxies are found in the central galaxies of the Virgo \citep{harris09} and Fornax galaxy clusters.  As population and dynamical
tracers GCs have a high significance for investigating the dark matter content and distribution as well as the population structure and formation
history of M87 and NGC 1399. 
Regarding NGC 1399, \citet{schuberth10} presented about 700 GC radial velocities within 80 kpc of galactocentric radius. Blue and red clusters  show distinct kinematical properties
with a sharp transition. The red clusters resemble the galaxy light, while blue clusters behave more erratic and are probably accreted.   
The best fitting dark halo agrees reasonably well with that from X-ray studies, but a substructure within the dark halo that
 has been suspected in earlier X-ray studies, has not been confirmed.   \\

M87 also received particular attention. \cite{strader11} provide a wealth of kinematical data for over 700 GCs
that \cite{agnello14} use for a dynamical analysis.  They found a dark matter fraction of 0.95 within a radius of 135 kpc and an inner dark matter profile that is steeper
than predicted by cosmological simulations. Such a big sample also permits to analyse the orbital properties of GCs. Remarkably, the anisotropy seems to be
mainly tangential which supports the idea that many objects on radial orbits have been dissolved. 
Near M87 has been found the object with the highest negative radial velocity detected so far, a GC with -1025 km/s (\citealt{caldwell14}). Its dynamical history
is mysterious.  \\

A significant increase of the database regarding metallicities \citep{usher12} and kinematics of GCs in early-type galaxies \citep{pota13} was achieved by the still ongoing "The SAGES Legacy Unifying Globulars and GalaxieS Survey (SLUGGS)"  \citep{brodie14}. Until now, almost 1000 metallicities and 2500 velocities of GCs in a dozen early-type galaxies have been published.  
Highlighting a few  galaxies: The disputed metallicity bimodality  of GCs in NGC 3115 has been confirmed  by \citet{brodie12} through CaII triplet strengths. Breaking the degeneracy in
dynamical models between potential and orbital anisotropy,  \citep{napolitano14} find the dark halo of NGC 5846 to be consistent
with cosmological simulations, that its stellar IMF is Salpeter-like, and that the GC orbits are isotropic in the central parts and slightly radial at larger distances \citep{napolitano14}.
More than 400 GCs with photometry  and velocities build the database for M60 \citep{pota15}. \\

The correlation between central supermassive black hole masses and properties of GCSs have been discussed in several papers  \citep{burkert10,sadoun12,pota13b,harrisg14}. The latter two
contributions could investigate larger galaxy samples and find   the correlations weaker than described before.  
Massive nuclear GCs  host supermassive black holes. \citet{graham09} provides a relation between the masses of a nuclear star cluster, the bulge and the black hole. 
\citet{antonini15} show how erosion by binary BHs may change the mass of a nuclear star cluster. \citet{georgiev14} provides a catalogue of 228 nuclear star clusters in nearby spirals. \\

The discussion regarding the characteristic parameters and the nature of UCDs is continuing. \citet{brodie11} find, on the basis of a survey in M87,
that the colour-magnitude relation of  UCDs is  offset from  that of GCs. They conclude that the majority of UCDs are stripped nuclei of former dwarf
galaxies.  \citet{norris11} point to the dual nature of UCDs characterised by mass. Above $7\times10^7 M_\odot$, UCDs are predominantly
stripped nuclei, while at lower masses, many "normal" GCs  may be mixed in.  
An interesting finding were the enhanced M/L-values of some UCDs in the Virgo cluster which could not be explained by stellar population models ,
perhaps indicative of dark matter \citep{hasegan05}. Later work on UCDs in the Fornax cluster did not confirm this \citep{hilker07}. In particular, 
 the brightest and resolved UCD in the Fornax cluster shows population properties consistent with existing models \cite{frank11}, but the discussion
 is going on \cite{mieske13}.
 The study of a large number of UCDs around M87 suggests that UCDs are distinct from GCs by their sizes rather than by their masses. The UCDs in 
 M87 show both radial distributions and  orbital properties different from those of GCs \citep{zhang15}.   
  \citet{bruens11} explain both compact and extended clusters by dynamical evolution in star cluster complexes. 
  Convincing evidence that some UCDs evolve from bigger parent galaxies, is the detection of a supermassive black hole 
in a  UCD in M60 that has 15\% of the mass of its host system \cite{seth14}. \\

Cosmological formation of GCs has not yet been identified, but GCs populating a galaxy cluster rather than an individual galaxy, have been 
found by \citet{peng11} in the Coma cluster,  \citet{west11} in Abell 1189, and by \citet{alamo13} in Abell 1689. For the latter cluster, the authors
quote a number of 160 000 GC within 400 kpc. 
The other extreme appears as "the most isolated globular clusters in the Local Universe"  in the
vicinities of M81 and M82 \citet{jang12}. The Local Group does not seem to host very isolated GCs \citep{ditullio13}.

\bibliography{comm37}

\end{document}